\documentclass[aps,prc,twocolumn,superscriptaddress,showpacs]{revtex4}
\usepackage{epsfig}

\usepackage{graphicx}
\usepackage{dcolumn}
\usepackage{bm}
\usepackage{ulem}
\usepackage{color}

\newcommand{\al}{$\alpha$}

\newcommand{\ran}{($\alpha$,n)}

\newcommand{\rng}{(n,$\gamma$)}

\newcommand{\rdp}{(d,p)}

\newcommand{\ovi}{$^{16}$O}
\newcommand{\ovii}{$^{17}$O}
\newcommand{\cano}{$^{13}$C($\alpha$,$n$)$^{16}$O}
\newcommand{\neanmg}{$^{22}$Ne($\alpha$,$n$)$^{25}$Mg}
\newcommand{\fdao}{$^{19}$F($d$,$\alpha$)$^{17}$O}
\newcommand{\ongo}{$^{16}$O($n$,$\gamma$)$^{17}$O}
\newcommand{\spro}{s-process}

\newcommand{\sfact}{S-factor}
\newcommand{\Nsv}{$N_A$$\left< \sigma v \right>$}

\begin{document}

\title{
  Broad levels in $^{17}$O and their relevance for the astrophysical
  s-process
}

\author{T.\ Faestermann}
\email{thomas.faestermann@ph.tum.de}
\affiliation{Physik Dept.~E12, Technische Universit\"at M\"unchen, D-85748 Garching, Germany}
\affiliation{Excellence Cluster Universe, D-85748 Garching, Germany}
\author{ P.\ Mohr}
\affiliation{
Diakonie-Klinikum, D-74523 Schw\"abisch Hall, Germany}
\affiliation{
Institute for Nuclear Research (ATOMKI), H-4001 Debrecen, Hungary}
 \author{R.\ Hertenberger}
\affiliation{Excellence Cluster Universe, D-85748 Garching, Germany}
\affiliation{Fakult\"at f\"ur Physik, Ludwig-Maximilians-Universit\"at M\"unchen, D-85748 Garching, Germany}
\author{H.-F.\ Wirth}
\affiliation{Fakult\"at f\"ur Physik, Ludwig-Maximilians-Universit\"at M\"unchen, D-85748 Garching, Germany}

\date{\today}

\begin{abstract}
Levels in $^{17}$O affect the astrophysical s-process in two opposite
ways. The neutron production is enhanced by resonances in the
$^{13}$C($\alpha$,$n$)$^{16}$O reaction at excitation energies around 7\,MeV
in $^{17}$O, and the number of available neutrons is reduced by low-lying resonances in
the $^{16}$O($n$,$\gamma$)$^{17}$O reaction corresponding to levels in
$^{17}$O with excitation energies of $4-5$\,MeV. The present work uses the
$^{19}$F($d$,$\alpha$)$^{17}$O reaction to determine absolute widths of the
relevant levels in $^{17}$O. The results improve the uncertainties of the
previously adopted values and resolve a discrepancy between recent studies for
the $1/2^+$ level close to the threshold of the $^{13}$C($\alpha$,$n$)$^{16}$O
reaction. In addition, improved excitation energies and widths are provided
for several states in $^{17}$O up to excitation energies close to 8\,MeV.
\end{abstract}

\pacs{25.45.-z, 25.55.Hp, 26.20.Kn}

\maketitle

\section{Introduction}
\label{sec:intro}
It is well-known that about one half of the nuclei heavier than iron are
synthesized in the astrophysical slow neutron capture process (\spro ). The
main component of the \spro\ is assigned to thermally pulsing AGB stars where
neutrons are generated by the \cano\ reaction at low temperatures ($kT \approx
8$\,keV) in the long interpulse phases and by the \neanmg\ reaction at higher
temperatures ($kT \approx 23$\,keV) in the shorter pulses. The weak component
of the \spro\ occurs in more massive stars with temperatures up to about $kT
\approx 90$\,keV \cite{The00,Kaepp06,Stra06,Kaepp11}. The role of levels in
\ovii\ for the \spro\ is twofold. First, levels close to and above the
$^{13}$C-\al\ threshold enhance the resonant neutron production in the
\cano\ reaction. Second, the number of neutrons available for capture
reactions on heavy nuclei is reduced by resonances in the \ongo\ reaction
which correspond to levels in \ovii\ close above the \ovi -$n$
threshold. Thus, the nucleus \ovi\ may act as a neutron poison via the
\ongo\ reaction. A detailed study of the role of $^{16}$O as neutron poison
for the \spro\ will be given elsewhere \cite{Mohr15}.

The present study attempts to provide improved level properties of states in
$^{17}$O. Although many experiments have been done over the last decades, the
adopted values for excitation energies $E^\ast$ and total widths $\Gamma$ are
often adopted in Ref. \cite{Til93} from early neutron scattering data
\cite{John73,Fow70}. These data affect the stellar reaction rates of the
\cano\ and \ongo\ reactions. In addition, the knowledge of these level
properties is essential for the analysis of indirect experimental data for the
\cano\ reaction rate. The \fdao\ reaction has been chosen for the present
study because practically all levels in \ovii\ are populated with sufficient
statistics. 

The paper is organized as follows. In Sect.~\ref{sec:exp} the experimental
procedure is briefly described, and the measured excitation energies $E^\ast$
and total widths are listed. Sect.~\ref{sec:c13an} focuses on the main
motivation of the present study which is the analysis of the $1/2^+$ state
very close to the $^{13}$C-\al\ threshold. Sect.~\ref{sec:o16ng}
presents the results for the two lowest resonances in the \ongo\ reaction
($3/2^-$, $E^\ast = 4.554$\,MeV and $3/2^+$, $E^\ast = 5.085$\,MeV). Some
interesting details for other levels will be discussed in
Sect.~\ref{sec:other}.

\section{Experimental procedure and results}
\label{sec:exp}
A first try to populate states in \ovii\ with the $^{16}$O(d,p) reaction had shown to produce too much background from the other components of the SiO$_2$ or Al$_2$O$_3$ targets.
The present work used the \fdao\ reaction to determine total widths of levels
in the residual \ovii\ nucleus. The pickup of a proton and a neutron is supposed to populate rather unspecifically all levels in the final nucleus. The experiment has been performed at the
MLL tandem accelerator of the Munich universities where a high-resolution Q3D magnetic spectrograph is
available. A deuteron beam was accelerated to an energy of 22 MeV and focused onto the target with an average intensity of more than 0.6 $\mu$A. As target we used $^6$LiF with a thickness of 46 $\mu$g/cm$^2$ evaporated onto a 12 $\mu$g/cm$^2$ carbon foil. The $^6$Li has the advantage that the (d,$\alpha$) reaction leads to another $\alpha$-particle with no excited states. The outgoing $\alpha$-particles were momentum analyzed with the Q3D spectrograph \cite{Loe77}. The identification and position measurement was performed with the 0.89 m long focal plane detector \cite{Wir01}. It consists of a proportional counter for energy loss and position measurement and a scintillator measuring the residual energy. $\alpha $ -spectra were taken with two settings: at a scattering angle of 15$^o$ and an excitation energy range between 3750 keV and 6200 keV and a long run at 10$^o$ and between 5500 and 7800 keV. Since the position along the focal plane is not a linear function of the particle energy we have used lines in \ovii\ for an internal calibration with a quadratic polynomial. And, since the slope of the calibration is not constant, the channel contents were accordingly transformed as well as their uncertainties.

\begin{figure}[tb]
\begin{center}
\begin{minipage}[t]{8.6 cm}
\epsfig{file=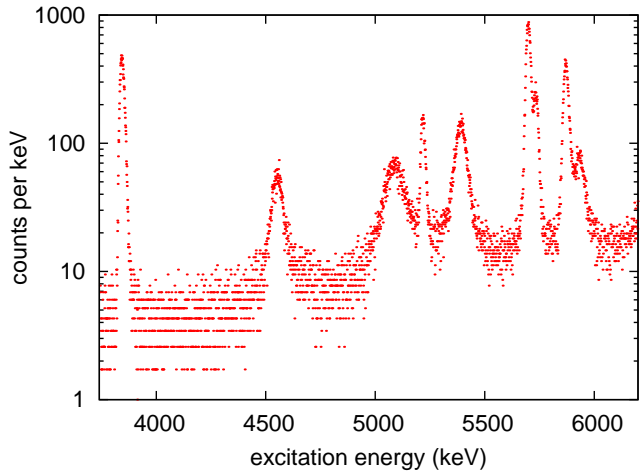,scale=0.52,angle=-90}
\end{minipage}
\begin{minipage}[t]{8.6 cm}
\caption{(Color online) Low energy spectrum for the 
$^{19}$F(d,$\alpha$)$^{17}$O reaction at $\Theta_{lab}$=15$^o$.
 \label{low}}
\end{minipage}
\end{center}
\end{figure}

\begin{figure}[tb]
\begin{center}
\begin{minipage}[t]{8.6 cm}
\epsfig{file=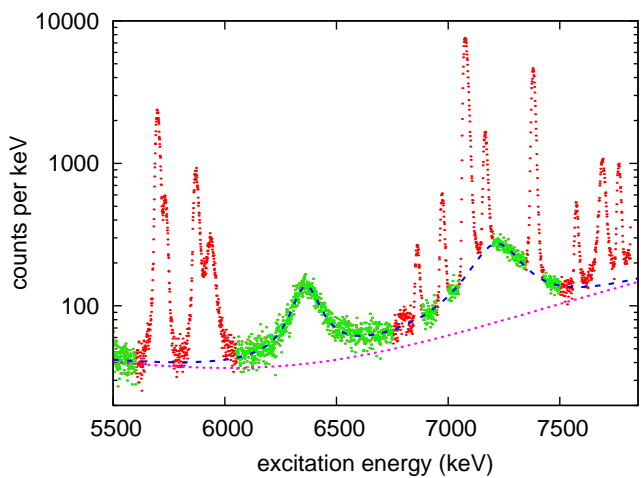,scale=0.52,angle=-90}
\end{minipage}
\begin{minipage}[t]{8.6 cm}
\caption{(Color online) High energy spectrum for the 
$^{19}$F(d,$\alpha$)$^{17}$O reaction at $\Theta_{lab}$=10$^o$. The data ranges that have been used to fit the two Lorentzians are indicated by the green (lighter) dots. The 
simultaneously fitted background (dotted) and the total fit function (dashed) are also shown. The strongest peak at 7075 keV is the only background line from the $^{16}$O(d,$\alpha_0$)$^{14}$N reaction.  \label{high}}
\end{minipage}
\end{center}
\end{figure}

The calibrated spectra are shown in Figs.\ \ref{low} and \ref{high}. All lines were fitted using a Gaussian or, for broad peaks, a Lorentzian line shape. The energy resolution of 20\,keV (FWHM) is mainly caused by the difference in energy loss of the 22 MeV deuterons and the 24.9 MeV (for $E^*$ = 6.36\,MeV) $\alpha$-particles. To fit the broad resonance at 6.36 MeV we used a large range of the spectrum from about 5.5 MeV to 7.5 MeV, excluding narrow peaks but including the broad peak at 7.20 MeV which extends down to the high energy tail of the 6.36 MeV line. For the background a cubic polynomial was used. Besides fitting with a constant width, the width of the 6.36~MeV line was also taken proportional to the velocity of the neutron, to be emitted, via the energy above the neutron-threshold at E$_{thr}$~=~4.143~MeV as 
\begin{equation}
\Gamma = \Gamma_0 \times \left[1+1/2(E-E_0)/(E_0-E_{thr})\right],
\label{eq:gam}
\end{equation}
\noindent
with $\Gamma_0$ the width at the resonance energy $E_0$. That improved the $\chi^2$ by about 1\%, reduced the resonance energy by 0.8 keV, but did not affect the width. The best fit has a $\chi^2$/dof~=~1344/1228. The resonance curve in fact is a Voigt-profile which is the convolution of a Gaussian and a Lorentzian and does not have an analytical solution. Therefore we have used the parametrization of Olivero and Longbothum \cite{Oli77} 
\begin{equation}
f_V\approx 0.5346*f_L+\sqrt{0.2166*f_L^2+f_G^2}
\label{eq:width}
\end{equation}
\noindent
to extract the Lorentzian width $f_L=\Gamma$ from the Voigt width $f_V$ (fitted) and the experimental Gaussian width $f_G \approx 20$~keV (FWHM). The results for the 6.36 MeV resonance as well as for all other states in $^{17}$O between 3.8 MeV and 7.8 MeV are given in Table \ref{tab:res}. To accommodate the uncertainty of the intrinsic energy calibration an uncertainty of 0.3~keV has been added in quadrature to the uncertainties of the fitted excitation energies.

\begin{table*}[htb]
\caption{
\label{tab:res}
Excitation energies $E^\ast$ and Lorentzian widths $\Gamma$  for states in $^{17}$O from this work compared to the adopted values \cite{Til93}. We have fitted Lorentzians (L) or Gaussians (G, \textit{width printed in italics}) resp. to the lines with widths FWHM, and deduced the intrinsic width $\Gamma$ for the Lorentzians as described in the text.
\\
}
\begin{tabular}{c r@{$\pm$}l r@{$\pm$}l c r@{$\pm$}l r@{$\pm$}l r@{$\pm$}l}
\multicolumn{1}{c}{$J^\pi$}
& \multicolumn{2}{c}{$E^\ast$ (keV) \cite{Til93}}
& \multicolumn{2}{c}{$\Gamma$ (keV) \cite{Til93}}
& \multicolumn{1}{c}{fit function}
& \multicolumn{2}{c}{$E^\ast$ (keV)\footnote{this work}} 
& \multicolumn{2}{c}{FWHM (keV)}
& \multicolumn{2}{c}{$\Gamma$ (keV)\footnotemark[1]}
 \\
\hline
$5/2^-$ 
& 3842.76 & 0.42 & \multicolumn{2}{c}{$\tau \le 25$\,fs} & G\footnote{used for energy calibration}
& 3842.9  & 0.4
& \textit{21.52} & \textit{0.21} \\
$3/2^-$
& 4553.8  &  1.6 & 40 & 5 & L\footnotemark[2]  
& 4551.4  &  0.7 & 48.2 & 1.7 
& 38.1 & 2.8\\
$3/2^+$
& 5084.8  &  0.9 & 96 & 5 & L 
& 5087.7  &  1.0 & 93.4 & 2.6 & 88 & 3 \\
$9/2^-$
& 5215.77 & 0.45 & \multicolumn{2}{c}{$< 0.1$} & G\footnotemark[2]  
& 5216.5  & 0.4 & \textit{21.6} & \textit{0.5} \\
$3/2^-$
& 5379.2  &  1.4 & 28 & 7 & L 
& 5388.8  &  0.6 & 49.4 & 1.1 & 39.0 & 2.1 \\
$7/2^-$
& 5697.26 & 0.33 & 3.4& 0.3 & G\footnotemark[2] 
& 5697.5  & 0.5& \textit{21.97} & \textit{0.14} \\
$(5/2^-)$
& 5732.79 & 0.52 & \multicolumn{2}{c}{$< 1$} & G\footnotemark[2]  
& 5731.6 & 0.4& \textit{21.97} & \textit{0.14}  \\
$3/2^+$
& 5869.07 & 0.55 & 6.6 & 0.7 & G\footnotemark[2]  
& 5869.7  & 0.6& \textit{25.2} & \textit{0.7} \\
$1/2^-$
& 5939   & 4   & 32 & 3 & L 
& 5931.0 & 1.1 & 44.7 & 3.0 & 33 & 5 \\
$1/2^+$
& 6356  & 8  & 124 & 12 & L 
& 6363.4 & 3.1 & 139 & 4 & 136 & 5 \\
$(5/2^+)$
& 6862 & 2  & \multicolumn{2}{c}{$< 1$} & G\footnotemark[2]
& 6860.7 & 0.4 & \textit{18.8} & \textit{0.7} \\
$(7/2^-)$
& 6972 & 2  & \multicolumn{2}{c}{$< 1$} & G\footnotemark[2]  
& 6972.6 & 0.4 & \textit{18.8} & \textit{0.4} \\
$5/2^-$
& 7165.7 & 0.8 & 1.38 & 0.05 & G\footnotemark[2]  
& 7165.4 & 1.8 & \textit{20.0} & \textit{0.5} \\
$3/2^+$
& 7202   & 10  & 280 & 30 & L 
& 7216 &  4  & 264 & 7 & 262 & 7 \\
$5/2^+$
& 7379.2 & 1.0 & 0.64 & 0.23 & G 
& 7380.1 & 0.4\footnote{not resolved}
& \textit{19.8} & \textit{0.5} \\
$5/2^-$
& 7382.2 & 1.0 & 0.96 & 0.20 & G 
& 7380.1 & 0.4\footnotemark[3] 
& \textit{19.8} & \textit{0.5} \\
$3/2^-$
& 7559 & 20 & 500 & 50  
&  \multicolumn{5}{c}{ } \\
$(7/2^+)$
& 7576 & 2  & \multicolumn{2}{c}{$< 0.1$} & G\footnotemark[2] 
& 7573.5 & 0.6 & \textit{18.4} & \textit{1.2} \\
$7/2^-$
& 7688.2 & 0.9 & 14.4 & 0.3 & L\footnotemark[2] 
& 7689.2 & 0.6 & 25.1 & 1.3 & 12 & 4 \\
$11/2^-$
& 7757 & 9 & \multicolumn{2}{c}{$-$} & G\footnotemark[2] 
& 7763.6 & 0.4 & \textit{18.1} & \textit{0.7} & \multicolumn{2}{c}{$< 4$} \\
\hline
\end{tabular}
\end{table*}

\section{The $1/2^+$ threshold level at $E^\ast = 6.36$\,MeV}
\label{sec:c13an}
We first focus on the neutron production in the \spro .  The \cano\ reaction
operates at very low temperatures of $kT \approx 8$\,keV; the Gamow window for
this temperature is located at $E \approx 190$\,keV; with the \al\ separation
energy $S_\alpha = 6358.69$\,keV \cite{Wang12} this corresponds to an excitation
energy of 
$E^\ast \approx 6550$\,keV. (Note that all energies $E$ are given in the
center-of-mass system except explicitly noted; excitation energies in
\ovii\ are denoted by $E^\ast$.) A detailed R-matrix study has taken into
account 84 levels in \ovii\ from $E^\ast = 0$ to $E^\ast \approx 20$\,MeV to
derive the astrophysical \sfact\ $S(E)$ and the reaction rate $N_A < \sigma v
>$ of the \cano\ reaction \cite{Heil08}. However, the careful inspection of
the level scheme of \ovii\ shows that the astrophysical \sfact\ in the Gamow
window is strongly affected by the properties of one particular broad $1/2^+$
state close 
to the \al\ threshold. The adopted parameters of this $1/2^+$ state have been
derived mainly from neutron scattering: $E^\ast = 6356 \pm 8$\,keV, $E =
-3$\,keV, $\Gamma = 124 \pm 12$\,keV \cite{Til93,John73,Fow70}. This level
will be called ``threshold level'' (TL) in the following.

The TL leads to a resonant (Breit-Wigner) contribution to the cross section
\begin{equation}
\sigma(E) = \frac{\pi}{k^2} \, 
\frac{\Gamma_\alpha(E) \, \Gamma_n(E+Q)}
     {(E-E_R)^2 + \Gamma^2/4}
\label{eq:BW}
\end{equation}
with the wave number $k$ and the energy-dependent widths $\Gamma_\alpha$ and
$\Gamma_n$ for the $\alpha$  and the neutron channel. The total width is
practically identical to the neutron width: $\Gamma \approx \Gamma_n$. The
spin factor $\omega = \frac{2J_R+1}{(2J_P+1)(2J_T+1)} = 1$ for this $1/2^+$
state has been omitted in
Eq.~(\ref{eq:BW}). Obviously, the cross section scales linearly with
$\Gamma_\alpha(E)$. In the Gamow window we have $E-E_R > \Gamma/2$, and thus
the cross section is also roughly proportional to $\Gamma_n(E+Q)$.

Because an adopted neutron width $\Gamma_n$ is available \cite{Til93}, most
recent work has focused on \al -transfer experiments and the indirect
determination of $\Gamma_\alpha(E)$ of the TL using spectroscopic factors,
reduced widths, or asymptotic normalization coefficients
\cite{Guo12,Pell08,John06,Kub03,Kee03,Avi15}. 
These studies have been complemented by a Trojan horse experiment
\cite{Cog13,Cog12}. A direct determination of $\Gamma_\alpha$ is impossible
for a subthreshold state and practically not possible for a state very close
above the threshold. Direct experimental data for the \cano\ reaction reach
energies down to about 270\,keV \cite{Dro93}. Further experimental data can be
found in Refs. \cite{Bru93,Har05,Heil08}; earlier experiments are summarized in the
NACRE compilations \cite{NACRE13,NACRE99}.

The present experiment improves the excitation energy $E^\ast$ and the total
width $\Gamma$ of the TL. Besides the direct impact on the cross section in
Eq.~(\ref{eq:BW}) and the resulting reaction rate, improved $E^\ast$ and
$\Gamma$ may also affect the analysis of the transfer experiments
\cite{Guo12,Pell08,John06,Kub03,Kee03}.  In all these studies the adopted
values for $E^\ast$ and $\Gamma$ of the TL \cite{Til93} had to be used to fit
small and broad peaks in spectra with significant background. Contrary to the
transfer experiments \cite{Guo12,Pell08,John06,Kub03,Kee03,Avi15}, the recent
Trojan horse experiment \cite{Cog13,Cog12} has attempted to derive $\Gamma$ and
$\Gamma_\alpha$ simultaneously; but also the Trojan horse experiment had to
use the adopted excitation energy $E^\ast$. Huge discrepancies of about a
factor of 30 for $\Gamma_\alpha$ and the contribution of the TL to the
\cano\ reaction rate have been derived from the transfer data and the Trojan
horse experiment
\cite{Guo12,Pell08,John06,Kub03,Kee03,Avi15,Cog13,Cog12}. This may 
at least partly be attributed to the use of the adopted values $E^\ast$ and
$\Gamma$ which are revised in the present study. Following the discussion in Ref. 
\cite{Avi15}, it has to be noted for completeness that the very low result of
\cite{John06} should be excluded.

Unfortunately, recent studies provide also discrepant results for the total
width $\Gamma$ in contradiction to the adopted values. The R-matrix fit by
Heil {\it et al.}\ \cite{Heil08} quotes $E^\ast = 6379.5$\,keV and $\Gamma =
158.1$\,keV, i.e.\ both values are larger and show an about 3$\sigma$
deviation from the adopted values. Contrary to the large values in the
R-matrix study \cite{Heil08}, the recent Trojan horse experiment claims a
smaller width of $\Gamma = 107 \pm 5_{\rm{stat}} {^{+9}_{-5}}_{\rm{norm}}$\,keV
\cite{Cog13}. A first analysis of these data has found an even smaller value
of $\Gamma = 83^{+9}_{-12}$\,keV \cite{Cog12}. 

The present results for this TL are $E^\ast = 6363.4 \pm 3.1$\,keV and $\Gamma =
136 \pm 5$\,keV. The new excitation energy $E^\ast$ is 7.4\,keV higher than
the adopted value. The uncertainty of $E^\ast$ has been reduced by more than a
factor of two. The new result for $E^\ast$ remains within $1\sigma$ of the
adopted value. The higher excitation energy changes this level from a
subthreshold level to a resonance at $E = 4.7 \pm 3$\,keV. The new width of
$\Gamma = 136 \pm 5$\,keV is 12\,keV higher than the adopted width, and it is
close to the average value of the high R-matrix result \cite{Heil08} and the
low Trojan-horse result \cite{Cog13}. The uncertainty of the width has been
improved significantly.


%
At first view it seems to be a simple task to estimate the impact of the
present new results on the $^{13}$C\ran $^{16}$O cross section and reaction
rate using Eq.~(\ref{eq:BW}). The following estimates are given for a typical
\spro\ temperature of $kT \approx 8$\,keV which corresponds to a most
effective energy $E \approx 190$\,keV. Keeping $\Gamma_\alpha$, the cross
section is enhanced by about 7\,\% from the increased new energy $E^\ast$ of
the TL, and the larger new total width $\Gamma$ leads to an increase of the
cross section by about 8\,\%. Combining both new values for $E^\ast$ and
$\Gamma$ increases the cross section by 15\,\%. However, this direct impact of
the new values has to be complemented by an indirect impact which is difficult
to quantify. As pointed out above, the experimental determination of
$\Gamma_\alpha(E)$ of the TL by indirect methods often requires a peak fitting
for the TL. In most cases these fits had to use the adopted values for
$E^\ast$ and $\Gamma$ because the corresponding peaks were very broad (and
sometimes located on non-negligible background). Thus, the present new results
for $E^\ast$ and $\Gamma$ of the TL should be used in a re-analysis of the
previous transfer and Trojan horse experiments to reduce the uncertainties of
the peak fitting procedures. This should lead to improved results for
$\Gamma_\alpha$ and the derived stellar reaction rate \Nsv ; however, such a
study must remain beyond the scope of the present work.

\section{The lowest resonances in the \ongo\ reaction}
\label{sec:o16ng}
At low energies the cross section of the \ongo\ capture reaction is dominated
by direct (non-resonant) $p$-wave capture to the $5/2^+$ ground state and
$1/2^+$ first excited state of $^{17}$O. The lowest resonances are found at
410.7\,keV ($3/2^-$) and 941.7\,keV ($3/2^+$). With $Q = 4143.08$\,keV
\cite{Wang12}, these energies correspond to $E^\ast = 4554$ and 5085\,keV.
 
The first resonance has an adopted width of $\Gamma = 40 \pm 5$\,keV. Because
of its spin $J = 3/2^-$, it does interfere with the direct capture amplitude,
and thus it affects the capture cross section down to about 250\,keV
\cite{Iga95,Ohs00,Mohr15}. The second resonance with $J^\pi = 3/2^+$ cannot
interfere with the direct $p$-wave capture. Although the adopted width of this
second resonance is large ($\Gamma = 96 \pm 5$\,keV), it has only very minor
influence on the stellar reaction rate of the \ongo\ reaction.

The present study provides total widths for the $3/2^-$ and $3/2^+$ states
which are close to the adopted values. We find $\Gamma = 38.1 \pm 2.8$\,keV
for the $3/2^-$ state and $88 \pm 3$\,keV for the $3/2^+$ state. Consequently,
the adopted reaction rate of the $^{16}$O\rng $^{17}$O reaction
\cite{KADONIS,Mohr15} does not change significantly from the slightly revised
widths of this study.

\section{Further results}
\label{sec:other}
For 12 levels in $^{17}$O the experimental width in the present study is given
by the resolution of the experiment. The lowest state of the present study
($5/2^-$, 3843\,keV) is located below particle thresholds, and thus a small
width is obvious. For five states upper limits below 1\,keV have been adopted;
the present work confirms that these levels are narrow. For five further
states small widths between 0.64\,keV and 6.6\,keV have been adopted. Again
the present study confirms these adopted values. No width is available in
\cite{Til93} for the $11/2^-$ state at $E^\ast = 7757$\,keV. The present work
is able to give an upper limit of about 4\,keV for the width, and we
determine a slightly higher value of $E^\ast = 7763.6$\,keV for the excitation
energy.

There are four further states in $^{17}$O with relatively broad widths which
could be determined in this work. For three of these levels our new results
for the widths are within $1\sigma$ of the adopted values \cite{Til93}. For
the relatively narrow ($\Gamma = 14.4 \pm 0.3$\,keV) $7/2^-$ level at
7689\,keV we find a slightly smaller width of $12 \pm 4$\,keV. However, as the
width of this level is smaller than our experimental resolution, the width
from this work has larger uncertainties than the adopted value. For the broad
$3/2^+$ level around $E^\ast \approx 7.2$\,MeV we find a slightly higher
$E^\ast = 7216$\,keV and a slightly smaller width $\Gamma = 262 \pm 7$\,keV
with a significantly reduced uncertainty compared to the adopted $\Gamma = 280
\pm 30$\,keV. The excitation energy of the $1/2^-$ state is reduced from
$E^\ast = 5939 \pm 4$\,keV to $5931.0 \pm 1.1$\,keV, and the adopted width of
$\Gamma = 32 \pm 3$\,keV is confirmed by the present result of $33 \pm
5$\,keV.

A surprising difference appears for the $3/2^-$ state with adopted $E^\ast =
5379.2 \pm 1.4$\,keV and $\Gamma = 28 \pm 7$\,keV. Here we find a higher
excitation energy $E^\ast = 5388.8 \pm 0.6$\,keV and a larger width $\Gamma =
39.0 \pm 2.1$\,keV. We do not have an explanation for the difference of the
excitation energy $E^\ast$ between the present work and the adopted value
which is based on the $^{16}$O\rdp $^{17}$O experiment by Piskor and
Sch\"aferlingova \cite{Pis90}. However, we note that the adopted width of
$\Gamma = 28 \pm 7$\,keV is based on an early $^{16}$O\rdp $^{17}$O experiment
by Browne \cite{Bro57}, whereas an early neutron scattering experiment by
Striebel {\it et al.}\ \cite{Str58} reports a much higher value of $\Gamma =
41.4$\,keV (without given uncertainty). 

It is interesting to note that an early experiment by Holt {\it et al.}\ \cite{Hol78} studied the $^{17}$O($\gamma$,n) reaction and deduced from the R-matrix analysis values for the width of five states in $^{17}$O which coincide with our values within our error bars, if inflated by 30$\%$. They unfortunately did not quote uncertainties. For the 6.36 MeV state they had a width of 130\,keV.

\section{Conclusions}
\label{sec:conc}
The present work has used the $^{19}$F(d,$\alpha$)$^{17}$O reaction to study
excitation energies $E^\ast$ and total widths $\Gamma$ of levels in the
$^{17}$O nucleus at excitaton energies between about 4 and 8\,MeV. Several
obtained widths have significantly smaller uncertainties than the adopted
values \cite{Til93}. The overall agreement with the adopted values
\cite{Til93} is good and remains typically within $1-2\ \sigma$ of the adopted
values.

The focus of the present study is the neutron production and absorption in the
astrophysical \spro . It is found that the role of $^{16}$O as a neutron
poison is not affected because the adopted widths of the first resonances in
the $^{16}$O\rng $^{17}$O reaction are essentially confirmed in this
work. The neutron production in the $^{13}$C\ran $^{16}$O reaction depends on
the properties of the $1/2^+$ threshold level. Contrary to the adopted value
of the excitation energy $E^\ast$, our new results show that this
threshold level is located a few keV above the $^{13}$C-\al\ threshold, and we
find a larger total width than adopted in \cite{Til93} with a significantly
reduced uncertainty.

\begin{acknowledgments}
We acknowledge gratefully the help of Paul Garrett (Univ. of Guelph) and his group with data taking and Vinzenz Bildstein (Guelph) for performing initial fits to the data. This work was supported by OTKA (K101328 and K108459) and the DFG cluster of excellence "Origin and Structure of the Universe" (www.universe-cluster.de).
\end{acknowledgments}


\vspace{1mm}

\noindent


\begin{thebibliography}{99}
%
\bibitem{The00}
L.-S.\ The, M.\ ElEid, B.\ S.\ Meyer,
\apj\ {\bf 533}, 998 (2000).
%
\bibitem{Kaepp06}
F.\ K\"appeler and A.\ Mengoni,
Nucl.\ Phys.\ {\bf 777}, 291 (2006).
%
\bibitem{Stra06}
O.\ Straniero, R.\ Gallino, S.\ Cristallo,
Nucl.\ Phys.\ {\bf 777}, 311 (2006).
%
\bibitem{Kaepp11}
F.\ K\"appeler, R.\ Gallino, S.\ Bisterzo, W.\ Aoki,
Rev.\ Mod.\ Phys.\ {\bf 83}, 157 (2011).
%
\bibitem{Mohr15}
P.\ Mohr, C.\ Heinz, A.\ Mengoni, I.\ Dillmann, M.\ Pignatari, F.\ K\"appeler,
\apj\, to be submitted.
%
\bibitem{Til93}
D.\ R.\ Tilley, H.\ R.\ Weller, C.\ M.\ Cheves,
Nucl.\ Phys.\ {\bf A564}, 1 (1993).
%
\bibitem{John73}
C.\ H.\ Johnson,
\prc\ {\bf 7}, 561 (1973); Erratum: \prc\ {\bf 8}, 851 (1973).
%
\bibitem{Fow70}
J.\ L.\ Fowler and C.\ H.\ Johnson,
\prc\ {\bf 2}, 124 (1970).
%
\bibitem{Loe77}
M.\ L\"offler, H.-J.\ Scheerer, H.\ Vonach,
Nucl.\ Instr.\ \& Meth.\ B {\bf 111}, 1 (1977).
%
\bibitem{Wir01}
H.-F. Wirth, PhD thesis, 
Technische Universit\"at M\"unchen, (2001).
\\
http://mediatum.ub.tum.de.eaccess.ub.tum.de
/node?id=602907
%
\bibitem{Oli77}
J.J. Olivero, R.L. Longbothum,
Journal of quantitative spectroscopy \& radiation transfer {\bf 17}, 233 (1977).
%
\bibitem{Wang12}
M.\ Wang, G.\ Audi, A.\ H.\ Wapstra, F.\ G.\ Kondev, M.\ MacCormick, X.\ Xu,
B.\ Pfeiffer,
Chin.\ Phys.\ C {\bf 36}, 1603 (2012).
%
\bibitem{Heil08}
M.\ Heil, R.\ Detwiler, R.\ E.\ Azuma, A.\ Couture, J.\ Daly, J.\ G\"orres,
F.\ K\"appeler, R.\ Reifarth, P.\ Tischhauser, C.\ Ugalde, M.\ Wiescher,
\prc\ {\bf 78}, 025803 (2008).
%
\bibitem{Guo12}
B.\ Guo {\it et al.},
\apj\ {\bf 756}, 193 (2012).
%
\bibitem{Pell08}
M.\ G.\ Pellegriti {\it et al.},
\prc\ {\bf 77}, 042801(R) (2008).
%
\bibitem{John06}
E.\ D.\ Johnson {\it et al.},
\prl\ {\bf 97}, 192701 (2006).
%
\bibitem{Kub03}
S.\ Kubono {\it et al.},
\prl\ {\bf 90}, 062501 (2003).
%
\bibitem{Kee03}
N.\ Keeley, K.\ W.\ Kemper, D.\ T.\ Khoa,
Nucl.\ Phys.\ {\bf A726}, 159 (2003).
%
\bibitem{Avi15}
M.\ L.\ Avila, G.\ V.\ Rogachev, E.\ Koshchiy, L.\ T.\ Baby, J.\ Belarge,
K.\ W.\ Kemper, A.\ N.\ Kuchera, D.\ Santiago-Gonzalez,
\prc\ {\bf 91}, 048801 (2015).
%
\bibitem{Cog13}
M.\ La Cognata {\it et al.},
\apj\ {\bf 777}, 143 (2013).
%
\bibitem{Cog12}
M.\ La Cognata {\it et al.},
\prl\ {\bf 109}, 232701 (2012).
%
\bibitem{Dro93}
H.\ W.\ Drotleff, A.\ Denker, H.\ Knee, M.\ Soin\'e, G.\ Wolf, J.\ W.\ Hammer,
U.\ Greife, C.\ Rolfs,, H.\ P.\ Trautvetter,
\apj\ {\bf 414}, 735 (1993).
%
\bibitem{Bru93}
C.\ R.\ Brune, I.\ Licot, R.\ W.\ Kavanagh,
\prc\ {\bf 48}, 3119 (1993).
%
\bibitem{Har05}
S.\ Harissopulos, H.\ W.\ Becker, J.\ W.\ Hammer, A.\ Lagoyannis, C.\ Rolfs,
F.\ Strieder,
\prc\ {\bf 72}, 062801(R) (2005).
%
\bibitem{NACRE13}
Y.\ Xu, K.\ Takahashi, S.\ Goriely, M.\ Arnould, M.\ Ohta, H.\ Utsunomiya,
Nucl.\ Phys.\ {\bf A918}, 61 (2013).
%
\bibitem{NACRE99} 
C.\ Angulo {\it et al.},
Nucl.\ Phys.\ {\bf A656}, 3 (1999). 
%
\bibitem{Iga95}
M.\ Igashira, Y.\ Nagai, K.\ Masuda, T.\ Ohsaki, H.\ Kitazawa,
\apj\ {\bf 441}, L89 (1995).
%
\bibitem{Ohs00}
T.\ Ohsaki {\it et al.},
Proc.\ 10$^{\rm{th}}$ Int.\ Symp.\ Capture Gamma-Ray Spectroscopy and Related
Topics, ed.\ S.\ Wender,
AIP Conf.\ Proc.\ {\bf{529}}, 458 (2000).
%
\bibitem{KADONIS}
KADoNiS data base, online at {\it www.kadonis.org}; \\
I.\ Dillmann, R.\ Plag, F.\ K\"appeler, T.\ Rauscher,
Proceedings of the Scientific Workshop on Neutron Measurements, Theory and
Applications, Geel, Belgium, April 28-30, 2009, Ed.\ F.-J.\ Hambsch, p.~65. 
%
\bibitem{Pis90}
S.\ Piskor and W.\ Sch\"aferlingova,
Nucl.\ Phys.\ {\bf A510}, 301 (1990).
%
\bibitem{Bro57}
C.\ P.\ Browne,
Phys.\ Rev.\ {\bf 108}, 1007 (1957).
%
\bibitem{Str58}
H.\ R.\ Striebel, S.\ E.\ Darden, W.\ Haeberli,
Nucl.\ Phys.\ {\bf 6}, 188 (1958).
%
\bibitem{Hol78}
R.\ J.\ Holt, H.\ E.\ Jackson, R.\ M.\ Laszewski, J.\ E.\ Monahan, and J.\ R.\ Specht,
\prc\ {\bf 18}, 1962 (1978).
%
\end{thebibliography}
\end{document}